\documentclass[%
aps,
prl,
superscriptaddress,
twocolumn,
showpacs,
amsmath,
amssymb,
amsfonts,
a4paper,
]{revtex4}

\usepackage{graphicx}
\usepackage{bm}
\usepackage{latexsym}
\usepackage{color}

\usepackage[overlay,absolute]{textpos}

\newcommand{\be}{\begin{equation}}
\newcommand{\ee}{\end{equation}}

\long\def\hide#1{}

\begin{document}

\title{
Scale invariant growth processes in expanding space
}

\author{Adnan Ali}
\affiliation{%
Centre for Complexity Science, University of Warwick,
Coventry CV4 7AL, United Kingdom
}
\author{Robin C. Ball}
\affiliation{%
Centre for Complexity Science, University of Warwick,
Coventry CV4 7AL, United Kingdom
}
\affiliation{%
Department of Physics, University of Warwick,
Coventry CV4 7AL, United Kingdom
}
\author{Stefan Grosskinsky}
\affiliation{%
Centre for Complexity Science, University of Warwick,
Coventry CV4 7AL, United Kingdom
}
\affiliation{%
Mathematics Institute, University of Warwick,
Coventry CV4 7AL, United Kingdom
}
\author{Ell\'ak Somfai}
\affiliation{%
Centre for Complexity Science, University of Warwick,
Coventry CV4 7AL, United Kingdom
}
\affiliation{%
Department of Physics, University of Warwick,
Coventry CV4 7AL, United Kingdom
}

\date{\today}

\begin{abstract}
Many growth processes lead to intriguing stochastic patterns and complex
fractal structures which exhibit local scale invariance properties. Such
structures can often be described effectively by space-time trajectories of
interacting particles, and their large scale behaviour depends on the overall
growth geometry. We establish an exact relation between statistical properties
of structures in uniformly expanding and fixed geometries, which preserves the
local scale invariance and is independent of other properties such as the
dimensionality. 
This relation generalizes standard conformal transformations as the natural
symmetry of self-affine growth processes.  We illustrate our main result
numerically for various structures of coalescing L\'evy flights and fractional
Brownian motions, including also branching and finite particle sizes.
One of the main benefits of this new approach
is a full, explicit description of the asymptotic statistics in
expanding domains, which are often non-trivial and random due to amplification
of initial fluctuations.
\end{abstract}

\maketitle

Scale invariant structures resulting from fractal growth processes are abundant
across nature \cite{mandelbrot-book-1982,meakin-1999}. Examples include
diffusion-limited aggregation
\cite{witten-prl-1981}, river basins \cite{rodriguez-iturbe-book-1997},
and self-affine domain boundaries forming behind growing fronts for spatial
competition models \cite{derrida-jpa-1991,saito-prl-1995}. While first results
appeared already 30 years ago, the field continues to be of interest
\cite{marchetti-2012} with recent applications in
microbial growth \cite{hallatschek-pnas-2007, korolev-rmp-2010}. In many cases
these structures can be modelled as trajectories of locally interacting
particles---a picture that we adopt in this Letter. The overall geometry has a
strong impact on growth processes. A dramatic
example is viscous fingering, where in constant width channel geometry a
stable Saffman-Taylor finger of fixed shape propagates
\cite{saffman-procrsoclonda-1958}, while in radial geometry a continuously
tip splitting branched structure emerges \cite{paterson-jfluidmech-1981,
thome-physfluidsa-1989}.
In biological growth spatial range expansion is often coupled to drift and
competition in the genetic pool \cite{hallatschek-evolution-2009}, and is 
recongnized to have major influence on the gene pool of natural
populations \cite{lehe-ploscomputbiol-2012}.

In this Letter we show how the effect of the overall geometry in many directed growth processes can be
captured elegantly in terms of a time dependent metric.  We view growing domain
boundaries as space-time trajectories of particles moving on the growth front,
which is expanding in many interesting cases.
A natural example within the scope of this Letter is isoradial growth in
two dimensions, 
such as domain boundaries of competing microbial species in a
Petri dish \cite{hallatschek-pnas-2007}.
While cosmology is an obvious example, there has been recent interest in
non-constant metric also in thin sheets \cite{klein-science-2007,
kim-science-2012, lee-prl-2012}.  Our results are applicable to the
formation of stochastic patterns and structures in a very general setting, 
including diffusion processes with
time-dependent diffusion rate (i.e., temperature) \cite{torre-pre-2000,
krapivsky-amjphys-2004, rador-pre-2006b, serino-jstatmech-2010},
in cosmologically expanding space \cite{herrmann-prd-2010}, or on a
biologically growing substrate.

In particular, we consider
self-affine space-time trajectories of particles under spatially
homogeneous but time dependent metric, and map those into more easily
tractable systems with constant metric. 
The mapping depends only on the local scale invariance exponent of the
trajectories, and works directly for local interactions which do not involve a
length scale, such as annihilation or coagulation of point particles.
Branching or exclusion/reflection of finite size particles can also be treated
after mapping the interaction length scales appropriately. 
This provides a natural extension of conformal maps to generalized self-affine
growth processes, and we show how this leads to an exact description of the
non-trivial asymptotic statistics of growth structures in expanding domains,
which is one particularly striking consequence of this new approach.

To describe our results in the most illustrative setting, we consider the
growth of self-affine structures (e.g., domain boundaries) in isoradial
geometry in two dimensions. These structures consist of directed ``arms'',
which can be interpreted as locally scale invariant space-time trajectories of
point particles moving in an expanding one dimensional space with periodic
boundary conditions. 

\begin{figure}[t]
\begin{center}
\includegraphics[bb=188 309 420 541,clip,width=37mm]{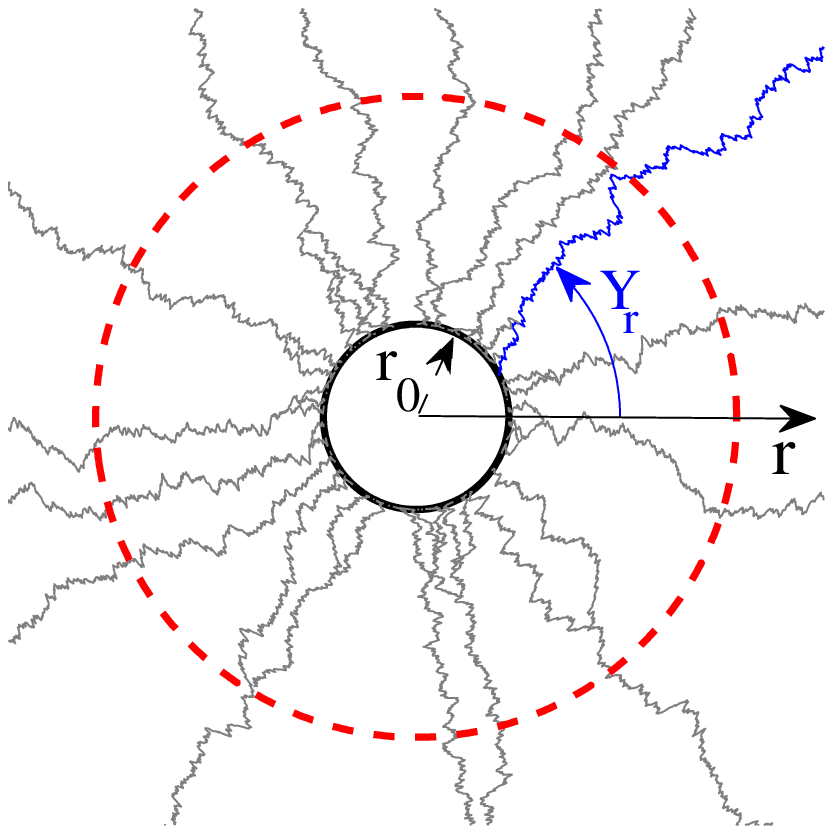}
\hspace{2mm}
\includegraphics[bb=163 287 470 554,clip,width=45mm]{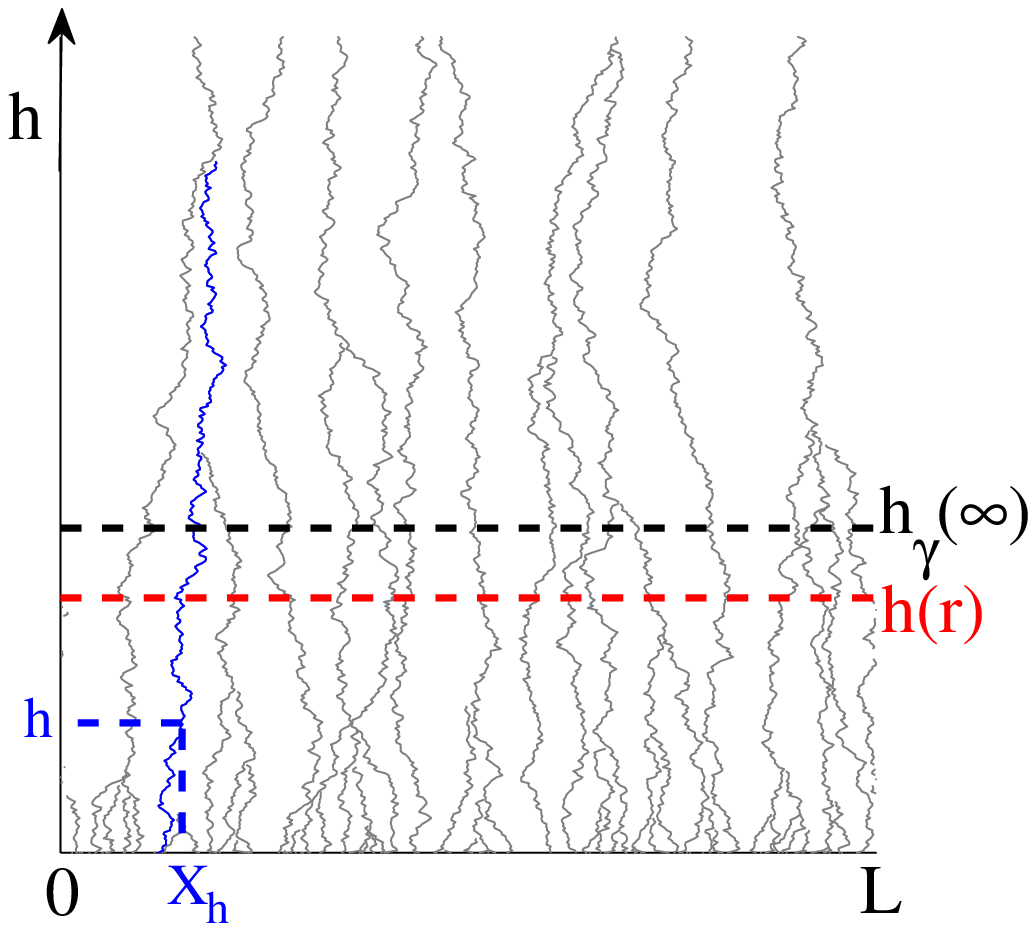}
\end{center}
\caption{\label{fig:radial}
(color online). Expanding radial growth structure and the same structure on
a fixed domain with periodic boundary conditions, illustrated for the case
of coalescing random walks ($\gamma =1/2$). The notation for the displacements $Y_r$ (\ref{yr}) and $X_h$ (\ref{xh}) is illustrated in blue. The distribution of the rescaled radial structure
at radius $r$ is identical to the distribution of the fixed domain structure
at height $h(r)$ as given by (\ref{mapping}), indicated by dashed red lines.
This mapping (plotted in Fig.~\ref{fig:mapping}) has a finite limit $h_\gamma (\infty )$ for $\gamma <1$, indicated by a dashed black line. Parameters are $L=100$ with $r_0 = L/2\pi$, 
unit diffusion coefficient and initially $100$ arms.
}
\end{figure}

Consider an isotropic radial structure growing from an initial disk with
radius $r_0$, shown in Fig.~\ref{fig:radial}(a) for an example of radial
coalescing Brownian motions, where also the following notation is illustrated.
We describe each arm by the displacement along the perimeter of the growing
circle
\be\label{yr}
  Y_r \in [0,2\pi r )\quad\mbox{with}\quad r\geq r_0
\ee
as a function of the radial distance $r$; directed radial growth means that
this is possible. 
In the increment 
\be\label{dY}
  dY_r = Y_r\,dr/r + d\tilde{Y}_r
\ee
the first term is 
due to the stretching of space, and the second corresponds to the
inherent fluctuations encoding the local scale invariance of the arms.
Instead of radial coordinates $(Y,r)$, the arms can also be represented in
modified polar coordinates $(X,h)$:  the polar angle is multiplied by $r_0$
and denoted by $X_h$, which is in a fixed periodic domain:
\be\label{xh}
  X_h \in [0,L)\quad\mbox{with}\quad h\geq 0 ,
\ee
and the relation between $r$ and $h$ will be determined shortly.
The choice $L=2\pi r_0$ enables matching the initial conditions between
$X_{h=0}$ and $Y_{r=r_0}$. 
This implies
\be
  X_h =\frac{r_0}{r}\, Y_r \,,
\ee
which using Eq.~(\ref{dY}) yields for the increments
\be
  dX_h = \frac{r_0}{r}d\tilde Y_r \,.
\ee
We impose that the mapping between
expanding and fixed geometry 
preserves the relevant local structure of the object (analogously
to conformal invariance), which in our case is
given by local scale invariance of the arms
\be\label{lsi}
  dX_h \sim (dh)^\gamma \quad \text{and} \quad d\tilde{Y}_r\sim (dr)^\gamma
\ee
with $\gamma >0$. For example, diffusive fluctuations correspond to $\gamma
=1/2$ (see \cite{ali-acs-2010} for related results), and for ballistic displacements of the arms $\gamma =1$. Other values
are related to sub- or superdiffusive behaviour, such as $\gamma =2/3$ for
domain boundaries driven by a surface in the KPZ universality class
\cite{saito-prl-1995,ferrari-pre-2006,ali-pre-2012}.

\begin{figure}[t]
\begin{center}
\includegraphics[width=0.8\columnwidth]{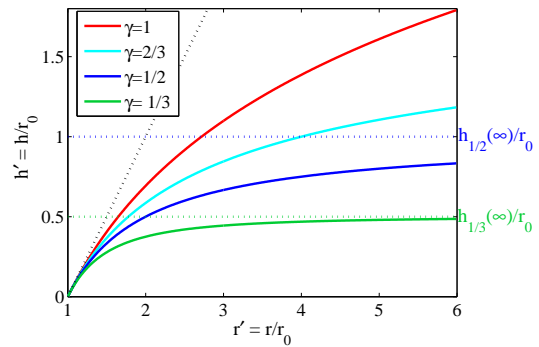}
\end{center}
\caption{\label{fig:mapping}
(color online). The mapping $h(r)$ between an expanding radial growth
structure and the same structure on a fixed domain as given in
(\ref{mapping}) in units of $r_0$, see (\ref{smap}).
Due to local equivalence of the two processes, 
$h(r)\approx r-r_0$ for $r\approx r_0$. The different
geometries affect the behaviour at large $r$, in particular $h(r)$ has a
finite limit $h_\gamma (\infty )$ for $\gamma <1$ (\ref{hlim}) and diverges for $\gamma\geq 1$. 
The asymptotic behaviour is
indicated by dotted lines, except for $\gamma=2/3$ which is off the figure.
}
\end{figure}

\begin{figure*}[t]
\begin{center}
\includegraphics[bb= 99 267 476 555,width=0.66\columnwidth]{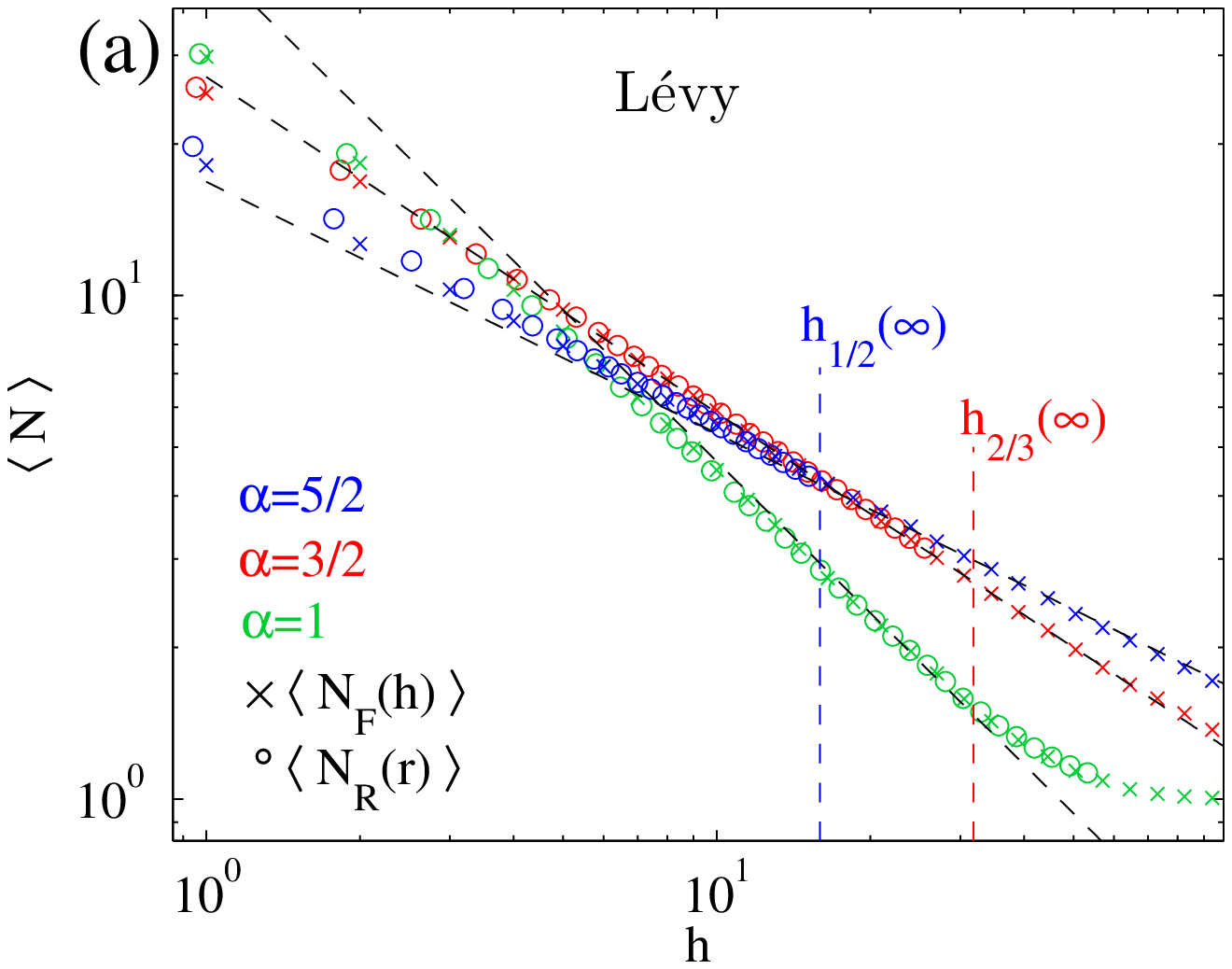}%
\hspace{3mm}%
\includegraphics[bb=116 279 461 544,width=0.66\columnwidth]{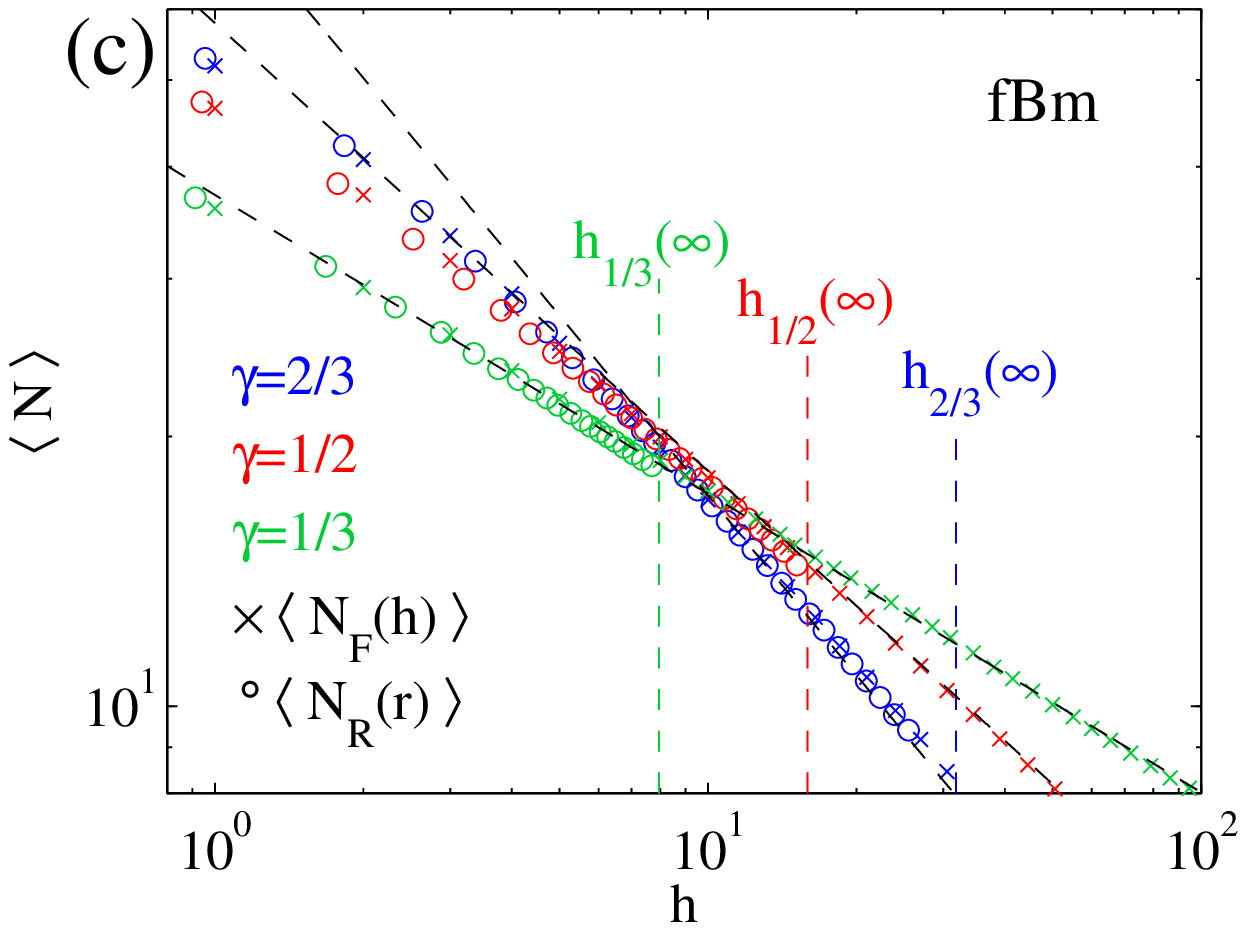}%
\hspace{3mm}%
\includegraphics[bb=116 279 461 544,width=0.66\columnwidth]{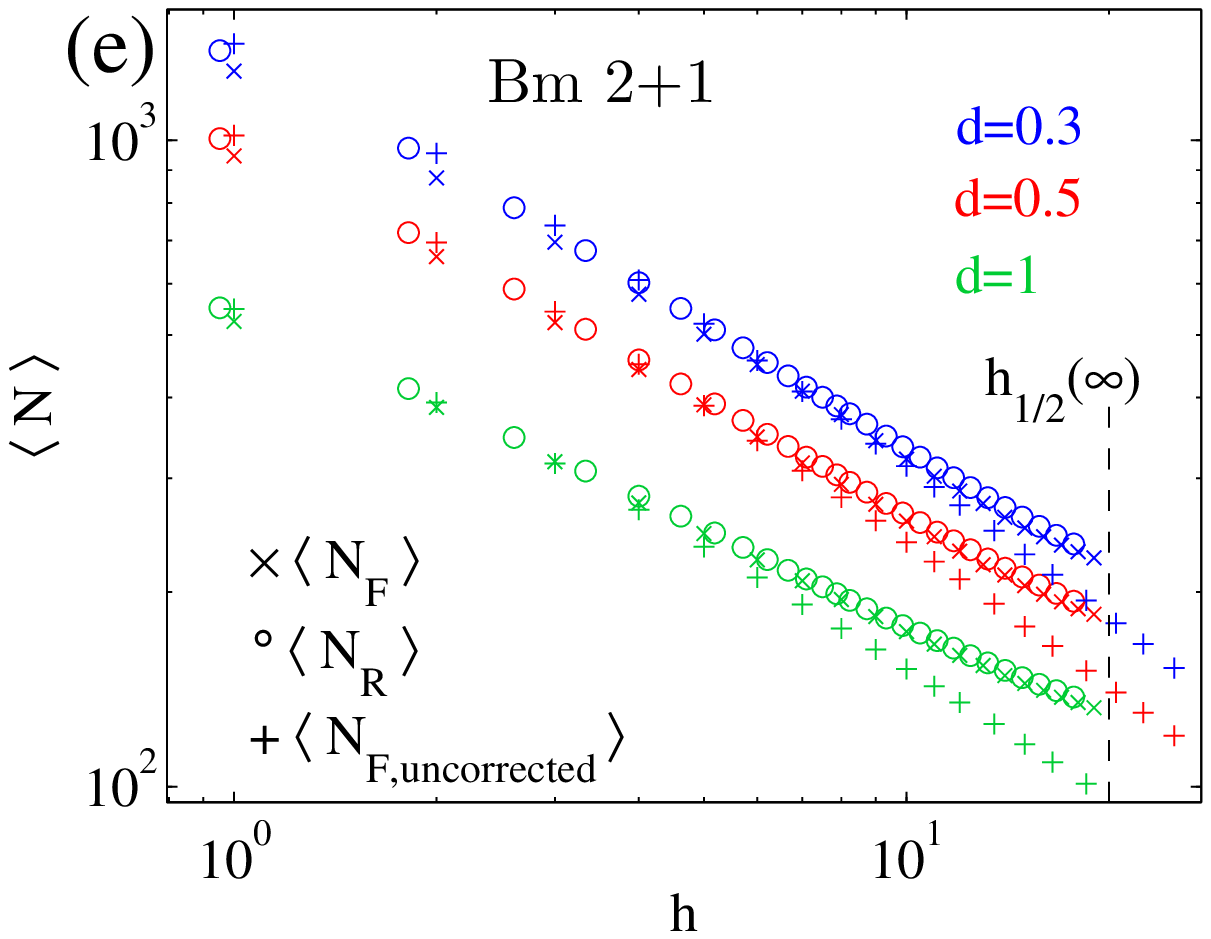}%
\vspace*{3mm}\newline%
\includegraphics[bb=102 267 476 558,width=0.66\columnwidth]{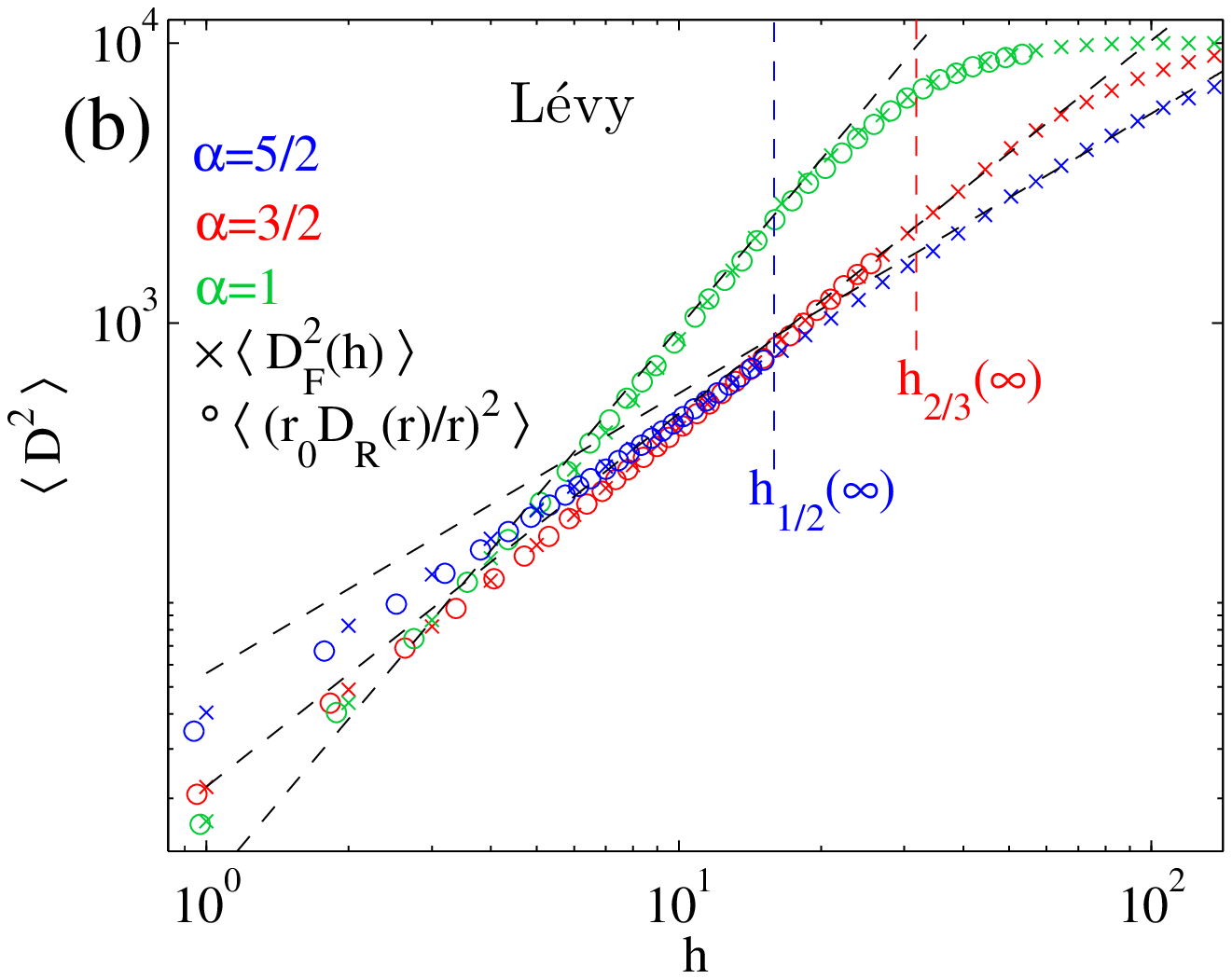}%
\hspace{3mm}%
\includegraphics[bb=115 284 461 544,width=0.66\columnwidth]{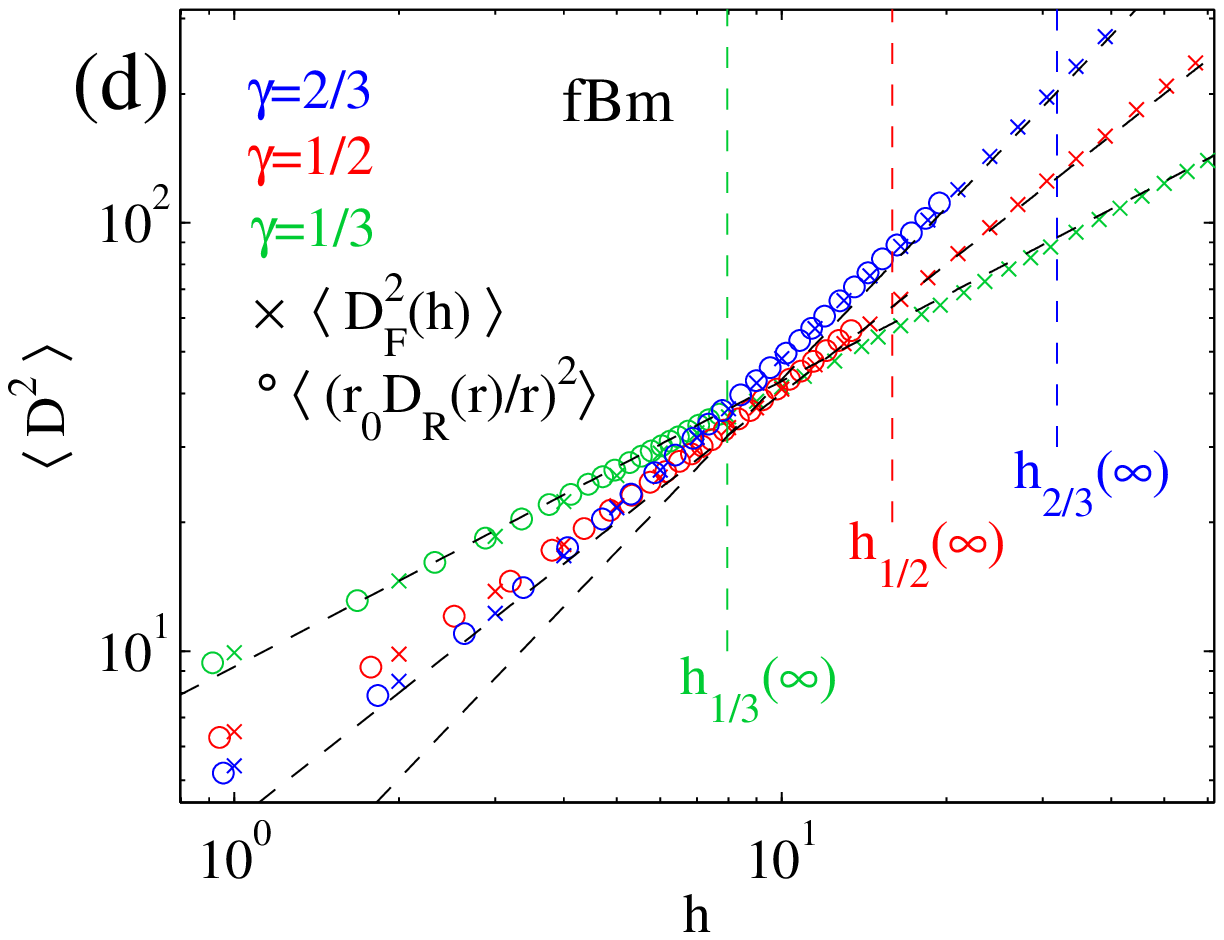}%
\hspace{3mm}%
\includegraphics[bb=116 279 461 544,width=0.66\columnwidth]{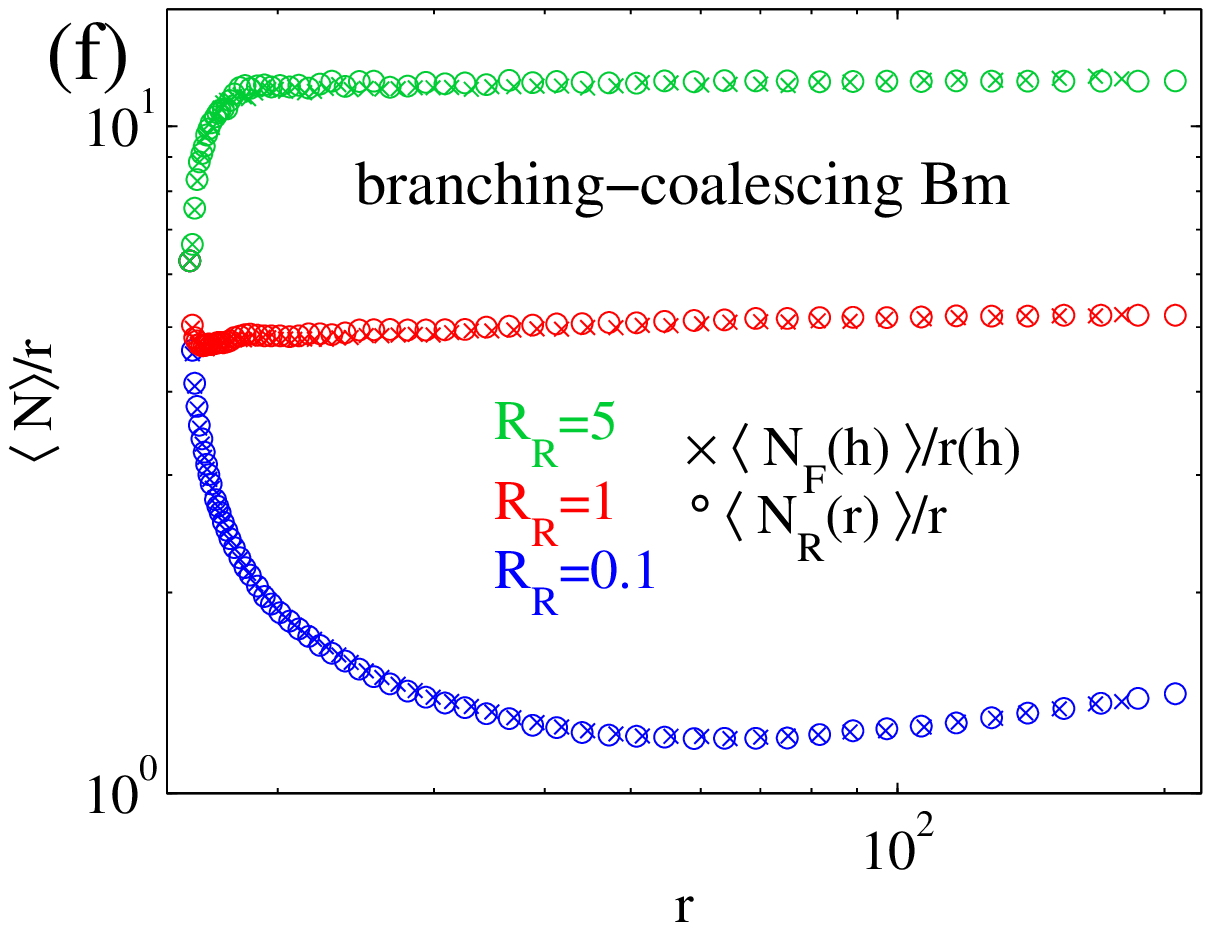}%
\end{center}
\caption{\label{fig:numerical}
(color online). Numerical demonstration of the mapping (\ref{mapping}) between
radial
geometry ($\circ$ on all panels) and fixed domain ($\times$):
(a), (b) number of surviving arms $\langle N\rangle$ and their root mean
square distance $\langle D^2\rangle$ (\ref{D}) for L\'evy flights (\ref{levy})
with $\gamma =\max\{ 1/\alpha ,1/2\}$;
(c), (d) same observables for fractional Brownian motion (\ref{fbm});
(e) number of surviving arms for Brownian motion ($\gamma =1/2$) with finite
particle size $d$, $+$ symbols indicate uncorrected, $\times$
corrected particle radii (see text);
(f) radial density of arms $\langle N\rangle/r$ for branching-coalescing
Brownian motion with fixed branching rate $R_R$ (\ref{rratio}) in the radial
geometry. 
For panels (a)-(e) the horizontal axis is $h$ or $h(r)$, with $h(\infty)$
indicated, while for panel (f) the horizonal axis is $r$ or $r(h)$.
The asymptotic scaling laws [black dashed lines on panels (a)-(d)]
break down when $\langle N\rangle\approx 1$.
Panels (a)-(d) and (f) are in $1+1$ dimensions, initially 100 arms, $L=100$
with $r_0=L/2\pi$, while (e) is in $2+1$ dimensions with $r_0=20$.
}
\end{figure*}

This leads to a relationship between $h\geq 0$ and $r\geq r_0$ via
\be
  \frac{dh}{dr}=\left(\frac{dX_h}{d\tilde{Y}_r}\right)^{1/\gamma}
  =\left(\frac{r_0}{r}\right)^{1/\gamma}\;,
\ee
where multiplicative prefactors, which are equal in (\ref{lsi})
although not indicated, drop out.
Integrating yields
\be\label{mapping}
h(r)=
\left\{\begin{array}{cl} r_0 \frac{\gamma}{1-\gamma}\left( 1-\left( r_0/r
\right)^{\frac{1-\gamma}{\gamma}}\right)\ &,\ \gamma\neq 1\\ 
r_0 \ln (r/r_0)\ &,\ \gamma =1\end{array} \right.
\ee
for all $r\geq r_0$.  For a single arm the matching initial condition
$Y_{r_0} =X_0$ leads to identical distributions 
$\frac{r_0}{r}\, Y_r \stackrel{\mathrm{dist.}}{=} X_{h(r)}$ for all $r\geq
r_0$.
Our main result is now that the same holds for the entire growth structures
which are characterized as collections of arms $\{ Y_r\}$ and $\{ X_h\}$,
with the independent variables linked through $h=h(r)$:
\be
\Big\{\frac{r_0}{r}\, Y_r \Big\}
\stackrel{\mathrm{dist.}}{=}\big\{ X_{h(r)} \big\} \quad
\text{for all }r\geq r_0 \;,
\ee
provided that the arms interact only locally. Examples of such
interactions include coagulation or annihilation, and we discuss
how this can be generalized in more detail below. Figure~\ref{fig:radial}
illustrates this correspondence for coalescing Brownian trajectories.

\emph{Properties of the mapping.}
To leading order $h(r)\approx r-r_0$ for $r$ close to $r_0$, since locally the
fixed domain and the radial models are equivalent. The effect of the different
geometries enters in the non-linear behaviour of $h(r)$ for larger values of
$r$, in particular for $0<\gamma<1$ we have
\be\label{hlim}
  h_\gamma (\infty )=\lim_{r\to\infty} h(r) =\frac{\gamma}{1-\gamma }\, r_0
  \;\;<\; \infty \,.
\ee
This observation is particularly interesting for coalescing or annihilating
structures, which exhibit an absorbing state in a fixed geometry with one or
no arms remaining as $h\to\infty$. Such structures often occur in neutral
models for competition in spatial populations \cite{saito-prl-1995,
hallatschek-pnas-2007}, and the absorbing state corresponds to fixation of the
model in one of the initial types.  By standard arguments the time to fixation
scales as $L^{1/\gamma} \sim r_0^{1/\gamma}$, which is much larger than
$h(\infty) \sim r_0$ for large systems.  For $\gamma <1$ we not
only confirm the previous (intuitive) result that there is no fixation in
expanding populations in the neutral case, but also give explicitly the
spatial distribution of the surviving types at large radii $r\to\infty$ as $\{
X_{h(\infty)}\}$. The process $X_h$ is much easier to simulate than $Y_r$, and in many cases
there also exist theoretical predictions \cite{followup}.

In Fig.~\ref{fig:mapping} we plot the mapping in convenient dimensionless
variables $r'=r/r_0$ and $h'=h/r_0$, so that
\be\label{smap}
  h'(r')=
  \left\{\begin{array}{cl}
    \frac{\gamma}{1-\gamma}\left( 1-\left( 1/r' \right)^{\frac{1-\gamma}{\gamma}
  }\right)\ &,\ \gamma\neq 1\\
  \log (r' )\ &,\ \gamma =1\end{array} \right.\ ,
\ee
for all $r'\geq 1$. For $\gamma =1$ we recover the generic conformal map from
the exterior of the unit circle to a strip, whereas for $\gamma\neq 1$
the mapping provides a natural generalization to self-affine processes. Note
that for general $\gamma$, $h'(r') =\log_q (r')$ is the $q$-logarithm with
$q=1/\gamma$ known from non-extensive statistical mechanics \cite{tsallis},
which can therefore also be interpreted as a generalization of conformal
transformations.

It is instructive to consider the mapping also for inward growing radial
structures, where $r\leq r_0$ (i.e., $r'\leq 1$), which formally leads to
negative heights $h<0$, corresponding to a fixed domain structure growing
downward. Observing the general relation
\be
  h' (1/r')=-(r')^{\frac{1-\gamma}{\gamma}}\, h' (r')\quad\text{for all }
  \gamma >0 \;,
\ee
all phenomena for such structures can be entirely understood by studying
outward growing ones. Note that in contrast to the expanding case now all
sub-ballistic structures lead to fixation since $|h'(r')|\to\infty$ as $r'\to
0$, whereas super-ballistic structures will have a non-trivial limit. First
results on inward growing radial structures have been obtained in
\cite{lebovka-jphysamathgen-1998}
and our approach provides a framework for a better understading of those which
is explained in detail in future work \cite{followup}.

\emph{Validity and locality.}
The mapping is based purely on a conservation of local scale invariance of the
structure. Therefore it is not surprising, that the mapping can be shown to
hold rigorously for processes which are fully determined by their local
structure, namely processes with independent increments such as Brownian
motion and self-similar L\'evy processes \cite{levyflights-book-1995}. On the other hand, there are
other self-similar processes with the same local scale invariance but more
complicated temporal correlations, such as fractional Brownian motion
(fBm) \cite{fbm-book-2010}. The correlations will influence the mapping and it
does not hold in general for such processes. Using fractional stochastic
calculus, one can derive a similar mapping for the particular model of fBm,
which leads to a more complex expression which is numerically very close to
(\ref{mapping}). This derivation is beyond the scope of this letter and is
discussed in detail in \cite{followup}.

In Fig.~\ref{fig:numerical} we illustrate the validity of the mapping for self
similar L\'evy flights, which are defined via independent stationary
increments with an $\alpha$-stable jump size distribution
\be
  \mathbb{P} (X_{h+}-X_h =x)\sim C|x|^{-(1+\alpha )}\qquad (\alpha >0)\;,
  \label{levy}
\ee
as well as fBm, which can be characterized as a
Gaussian process with covariances
\be\label{fbm}
  \left\langle X_{h+\Delta h}X_h \right\rangle
  \sim (h+\Delta h)^{2\gamma} + h^{2\gamma} -
  (\Delta h)^{2\gamma}\;.
\ee
L\'evy flights have local scale invariance parameter $\gamma =\max\{1/\alpha
,1/2\}$. They are super-diffusive and have non-continuous paths for $\alpha
<2$, and scale diffusively for $\alpha >2$ where the jump size has finite
variance.  fBm can be super- or sub-diffusive and is not Markovian, but still
the mapping (\ref{mapping}) works very well also in that case. In
Fig.~\ref{fig:numerical} we compare two statistics for coalescing interaction:
the average number of arms $\langle N\rangle$, and the total mean squared
distance between neighboring arms, as a measure for their spatial
distribution. For fixed geometry
\be
  D_F (h)^2 =\sum_{i=1}^{N(h)} \left(X_h^{(i+1)} -X_h^{(i)} \right)^2 \;,
  \label{D}
\ee
with an analogous $D_R (r)$ for radial geometry. Plotting the fixed and
circular data against $h$ and $h(r)$, respectively, we obtain a data collapse.
The power-law predictions for the fixed system in panels (a) to (d) can be
derived easily by standard mean-field arguments
\cite{alemany-1995,zaboronski-2006}.

A natural step to include non-local interactions is to introduce a particle
size. For simplicity we consider isotropic shapes with
diameter $d$, i.e., particles coagulate or annihilate already at a non-zero
distance $d$. As long as the diameter is much smaller than the macroscopic
length scale in the system, $d\ll r_0$, the corrections introduced are small.
Still they can be taken into account exactly by comparing the radial system with the
fixed domain one, where the particle diameter decreases as
$\frac{r_0}{r(h)}d$. In Fig.~\ref{fig:numerical}(e) we show both cases,
with and without this correction, for coalescing Brownian motions. We see that for small $d$ the
mapping still works very well even without corrections. Unlike all other
numerical data presented in this paper, this one is for an expanding sphere in
$2+1$ dimensions.
Finite range interactions are particularly important in higher dimensions, where 
coalescence or annihilation of point particles does not strictly occur,
they only get arbitrarily close to each other. The mapping is independent of
the dimension, as discussed below.

Another natural interaction included in growing structures is branching. This
is not a purely geometric interaction but has its own characteristic rate $R$,
which introduces a time scale in the system. For the mapped processes to have
the same statistics we require that the number of branching events $\Delta_F (dh)$
in the fixed domain model during a time interval $dh$ is the same as $\Delta_R
(dr)$ for the corresponding radial system. This implies a relation between
the branching rates
\be\label{rratio}
  \frac{R_R}{R_F}=\frac{\Delta_R (dr)/dr}{\Delta_F (dh)/dh}=\frac{dh}{dr} \;,
\ee
which is $(r_0/r)^{1/\gamma}$.
Thus to understand the density of branches $N_R (r)/r$ in a radially growing
system with fixed branching rate $R_R$, one has to compare to a fixed domain
system with increasing branching rate $R_F (h)=R_R\, r(h)^2 /r_0^2$, where
$r(h)$ is the inverse of (\ref{mapping}). Note that this rate diverges as
$r\to\infty$ or $h\to h(\infty)$. The density of branches for three different
branching rates is shown in Fig.~\ref{fig:numerical}(f) for Brownian
motions with $\gamma =1/2$.

\emph{Generalized geometries.}
Our results can be directly generalized to an arbitrary time dependent domain
of size $L(t)$ with homogeneous metric. We obtain
\be\label{tmap}
  h(t)=\int_0^t \Big(\frac{L(0)}{L(s)}\Big)^{1/\gamma} \, ds
\ee
analogously to Eq.~(\ref{mapping}). For example one can study exponentially
increasing domains, which is analogous to structures with exponentially
decreasing diffusivity. These have been studied in detail for single random
walks \cite{torre-pre-2000, krapivsky-amjphys-2004, rador-pre-2006b,
serino-jstatmech-2010} and are used in simulated annealing
\cite{kirkpatrick-science-1983}.

In $n+1$ dimensions, where $n$ is the spatial dimensionality, our method
applies directly if the scale invariance holds in all spatial directions
$i=1,\ldots ,n$
\be\label{dxi}
  dX_i \sim (dh)^\gamma \quad\text{and}\quad d\tilde Y_i\sim (dr)^\gamma \,.
\ee
It is possible to have anisotropy (possible $i$-dependence of the
multiplicative factors which are not indicated), 
but $\gamma$ should be identical in all directions. Then the mapping
(\ref{tmap}) stays exactly the same.

\emph{Summary.} We have demonstrated that a large class of locally
scale invariant, directed complex structures growing in radial or general
increasing geometries can be mapped to structures in fixed domains, which are
simpler and for which exact results are often available. This approach
provides an elegant and remarkably simple way to understand various phenomena related to
time-dependent metric, such as the effect of range expansions in competitive
biological growth. A particularly striking example is a full description of
the limiting statistics of radial competition interfaces.
Further examples and technical aspects are
discussed in more detail in \cite{followup}.

\begin{acknowledgments}
This work was supported by the Engineering and Physical Sciences Research
Council (EPSRC), Grant No. EP/E501311/1.
\end{acknowledgments}


\begin{thebibliography}{33}
\expandafter\ifx\csname natexlab\endcsname\relax\def\natexlab#1{#1}\fi
\expandafter\ifx\csname bibnamefont\endcsname\relax
  \def\bibnamefont#1{#1}\fi
\expandafter\ifx\csname bibfnamefont\endcsname\relax
  \def\bibfnamefont#1{#1}\fi
\expandafter\ifx\csname citenamefont\endcsname\relax
  \def\citenamefont#1{#1}\fi
\expandafter\ifx\csname url\endcsname\relax
  \def\url#1{\texttt{#1}}\fi
\expandafter\ifx\csname urlprefix\endcsname\relax\def\urlprefix{URL }\fi
\providecommand{\bibinfo}[2]{#2}
\providecommand{\eprint}[2][]{\url{#2}}

\bibitem[{\citenamefont{Mandelbrot}(1982)}]{mandelbrot-book-1982}
\bibinfo{author}{\bibfnamefont{B.~B.} \bibnamefont{Mandelbrot}},
  \emph{\bibinfo{title}{The Fractal Geometry of Nature}}
  (\bibinfo{publisher}{W. H. Freeman and Company}, \bibinfo{year}{1982}).

\bibitem[{\citenamefont{Meakin}(1999)}]{meakin-1999}
\bibinfo{author}{\bibfnamefont{P.}~\bibnamefont{Meakin}}, \bibinfo{journal}{J.
  Sol-Gel Sci. Technol.} \textbf{\bibinfo{volume}{15}}, \bibinfo{pages}{97}
  (\bibinfo{year}{1999}).

\bibitem[{\citenamefont{Witten and Sander}(1981)}]{witten-prl-1981}
\bibinfo{author}{\bibfnamefont{T.~A.} \bibnamefont{Witten}} \bibnamefont{and}
  \bibinfo{author}{\bibfnamefont{L.~M.} \bibnamefont{Sander}},
  \bibinfo{journal}{Phys.\ Rev.\ Lett.} \textbf{\bibinfo{volume}{47}},
  \bibinfo{pages}{1400} (\bibinfo{year}{1981}).

\bibitem[{\citenamefont{Rodriguez-Iturbe and
  Rinaldo}(1997)}]{rodriguez-iturbe-book-1997}
\bibinfo{author}{\bibfnamefont{I.}~\bibnamefont{Rodriguez-Iturbe}}
  \bibnamefont{and} \bibinfo{author}{\bibfnamefont{A.}~\bibnamefont{Rinaldo}},
  \emph{\bibinfo{title}{Fractal River Basins}} (\bibinfo{publisher}{Cambridge
  University Press}, \bibinfo{address}{Cambridge}, \bibinfo{year}{1997}).

\bibitem[{\citenamefont{Derrida and Dickman}(1991)}]{derrida-jpa-1991}
\bibinfo{author}{\bibfnamefont{B.}~\bibnamefont{Derrida}} \bibnamefont{and}
  \bibinfo{author}{\bibfnamefont{R.}~\bibnamefont{Dickman}},
  \bibinfo{journal}{J.\ Phys.\ A: Math.\ Gen.} \textbf{\bibinfo{volume}{24}},
  \bibinfo{pages}{L191} (\bibinfo{year}{1991}).

\bibitem[{\citenamefont{Saito and M{\"u}ller-Krumbhaar}(1995)}]{saito-prl-1995}
\bibinfo{author}{\bibfnamefont{Y.}~\bibnamefont{Saito}} \bibnamefont{and}
  \bibinfo{author}{\bibfnamefont{H.}~\bibnamefont{M{\"u}ller-Krumbhaar}},
  \bibinfo{journal}{Phys.\ Rev.\ Lett.} \textbf{\bibinfo{volume}{74}},
  \bibinfo{pages}{4325} (\bibinfo{year}{1995}).

\bibitem[{\citenamefont{Marchetti et~al.}(2012)\citenamefont{Marchetti, Taloni,
  Caglioti, Loreto, and Pietronero}}]{marchetti-2012}
\bibinfo{author}{\bibfnamefont{R.}~\bibnamefont{Marchetti}},
  \bibinfo{author}{\bibfnamefont{A.}~\bibnamefont{Taloni}},
  \bibinfo{author}{\bibfnamefont{E.}~\bibnamefont{Caglioti}},
  \bibinfo{author}{\bibfnamefont{V.}~\bibnamefont{Loreto}}, \bibnamefont{and}
  \bibinfo{author}{\bibfnamefont{L.}~\bibnamefont{Pietronero}},
  \bibinfo{journal}{Phys. Rev. Lett.} \textbf{\bibinfo{volume}{109}},
  \bibinfo{pages}{065501} (\bibinfo{year}{2012}).

\bibitem[{\citenamefont{Hallatschek et~al.}(2007)\citenamefont{Hallatschek,
  Hersen, Ramanathan, and Nelson}}]{hallatschek-pnas-2007}
\bibinfo{author}{\bibfnamefont{O.}~\bibnamefont{Hallatschek}},
  \bibinfo{author}{\bibfnamefont{P.}~\bibnamefont{Hersen}},
  \bibinfo{author}{\bibfnamefont{S.}~\bibnamefont{Ramanathan}},
  \bibnamefont{and} \bibinfo{author}{\bibfnamefont{D.~R.}
  \bibnamefont{Nelson}}, \bibinfo{journal}{Proc.\ Natl.\ Acad.\ Sci.\ USA}
  \textbf{\bibinfo{volume}{104}}, \bibinfo{pages}{19926}
  (\bibinfo{year}{2007}).

\bibitem[{\citenamefont{Korolev et~al.}(2010)\citenamefont{Korolev, Avlund,
  Hallatschek, and Nelson}}]{korolev-rmp-2010}
\bibinfo{author}{\bibfnamefont{K.~S.} \bibnamefont{Korolev}},
  \bibinfo{author}{\bibfnamefont{M.}~\bibnamefont{Avlund}},
  \bibinfo{author}{\bibfnamefont{O.}~\bibnamefont{Hallatschek}},
  \bibnamefont{and} \bibinfo{author}{\bibfnamefont{D.~R.}
  \bibnamefont{Nelson}}, \bibinfo{journal}{Rev.\ Mod.\ Phys.}
  \textbf{\bibinfo{volume}{82}}, \bibinfo{pages}{1691} (\bibinfo{year}{2010}).

\bibitem[{\citenamefont{Saffman and Taylor}(1958)}]{saffman-procrsoclonda-1958}
\bibinfo{author}{\bibfnamefont{P.~G.} \bibnamefont{Saffman}} \bibnamefont{and}
  \bibinfo{author}{\bibfnamefont{G.}~\bibnamefont{Taylor}},
  \bibinfo{journal}{Proc.\ R.\ Soc.\ Lond.\ A} \textbf{\bibinfo{volume}{245}},
  \bibinfo{pages}{312} (\bibinfo{year}{1958}).

\bibitem[{\citenamefont{Paterson}(1981)}]{paterson-jfluidmech-1981}
\bibinfo{author}{\bibfnamefont{L.}~\bibnamefont{Paterson}},
  \bibinfo{journal}{J.\ Fluid Mech.} \textbf{\bibinfo{volume}{113}},
  \bibinfo{pages}{513} (\bibinfo{year}{1981}).

\bibitem[{\citenamefont{Thome et~al.}(1989)\citenamefont{Thome, Rabaud, Hakim,
  and Couder}}]{thome-physfluidsa-1989}
\bibinfo{author}{\bibfnamefont{H.}~\bibnamefont{Thome}},
  \bibinfo{author}{\bibfnamefont{M.}~\bibnamefont{Rabaud}},
  \bibinfo{author}{\bibfnamefont{V.}~\bibnamefont{Hakim}}, \bibnamefont{and}
  \bibinfo{author}{\bibfnamefont{Y.}~\bibnamefont{Couder}},
  \bibinfo{journal}{Phys.\ Fluids A} \textbf{\bibinfo{volume}{1}},
  \bibinfo{pages}{224} (\bibinfo{year}{1989}).

\bibitem[{\citenamefont{Hallatschek and
  Nelson}(2009)}]{hallatschek-evolution-2009}
\bibinfo{author}{\bibfnamefont{O.}~\bibnamefont{Hallatschek}} \bibnamefont{and}
  \bibinfo{author}{\bibfnamefont{D.~R.} \bibnamefont{Nelson}},
  \bibinfo{journal}{Evolution} \textbf{\bibinfo{volume}{64}},
  \bibinfo{pages}{193} (\bibinfo{year}{2009}).

\bibitem[{\citenamefont{Lehe et~al.}(2012)\citenamefont{Lehe, Hallatschek, and
  Peliti}}]{lehe-ploscomputbiol-2012}
\bibinfo{author}{\bibfnamefont{R.}~\bibnamefont{Lehe}},
  \bibinfo{author}{\bibfnamefont{O.}~\bibnamefont{Hallatschek}},
  \bibnamefont{and} \bibinfo{author}{\bibfnamefont{L.}~\bibnamefont{Peliti}},
  \bibinfo{journal}{PLoS Comput.\ Biol.} \textbf{\bibinfo{volume}{8}},
  \bibinfo{pages}{e1002447} (\bibinfo{year}{2012}).

\bibitem[{\citenamefont{Klein et~al.}(2007)\citenamefont{Klein, Efrati, and
  Sharon}}]{klein-science-2007}
\bibinfo{author}{\bibfnamefont{Y.}~\bibnamefont{Klein}},
  \bibinfo{author}{\bibfnamefont{E.}~\bibnamefont{Efrati}}, \bibnamefont{and}
  \bibinfo{author}{\bibfnamefont{E.}~\bibnamefont{Sharon}},
  \bibinfo{journal}{Science} \textbf{\bibinfo{volume}{315}},
  \bibinfo{pages}{1116} (\bibinfo{year}{2007}).

\bibitem[{\citenamefont{Kim et~al.}(2012)\citenamefont{Kim, Hanna, Byun,
  Santangelo, and Hayward}}]{kim-science-2012}
\bibinfo{author}{\bibfnamefont{J.}~\bibnamefont{Kim}},
  \bibinfo{author}{\bibfnamefont{J.~A.} \bibnamefont{Hanna}},
  \bibinfo{author}{\bibfnamefont{M.}~\bibnamefont{Byun}},
  \bibinfo{author}{\bibfnamefont{C.~D.} \bibnamefont{Santangelo}},
  \bibnamefont{and} \bibinfo{author}{\bibfnamefont{R.~C.}
  \bibnamefont{Hayward}}, \bibinfo{journal}{Science}
  \textbf{\bibinfo{volume}{335}}, \bibinfo{pages}{1201} (\bibinfo{year}{2012}).

\bibitem[{\citenamefont{Lee et~al.}(2012)\citenamefont{Lee, Zhang, Jiang, and
  Fang}}]{lee-prl-2012}
\bibinfo{author}{\bibfnamefont{H.}~\bibnamefont{Lee}},
  \bibinfo{author}{\bibfnamefont{J.}~\bibnamefont{Zhang}},
  \bibinfo{author}{\bibfnamefont{H.}~\bibnamefont{Jiang}}, \bibnamefont{and}
  \bibinfo{author}{\bibfnamefont{N.~X.} \bibnamefont{Fang}},
  \bibinfo{journal}{Phys.\ Rev.\ Lett.} \textbf{\bibinfo{volume}{108}},
  \bibinfo{pages}{214304} (\bibinfo{year}{2012}).

\bibitem[{\citenamefont{de~la Torre et~al.}(2000)\citenamefont{de~la Torre,
  Maltz, M{\'a}rtin, Catuogno, and Garc{\'\i}a-Mata}}]{torre-pre-2000}
\bibinfo{author}{\bibfnamefont{A.~C.} \bibnamefont{de~la Torre}},
  \bibinfo{author}{\bibfnamefont{A.}~\bibnamefont{Maltz}},
  \bibinfo{author}{\bibfnamefont{H.~O.} \bibnamefont{M{\'a}rtin}},
  \bibinfo{author}{\bibfnamefont{P.}~\bibnamefont{Catuogno}}, \bibnamefont{and}
  \bibinfo{author}{\bibfnamefont{I.}~\bibnamefont{Garc{\'\i}a-Mata}},
  \bibinfo{journal}{Phys.\ Rev.\ E} \textbf{\bibinfo{volume}{62}},
  \bibinfo{pages}{7748} (\bibinfo{year}{2000}).

\bibitem[{\citenamefont{Krapivsky and Redner}(2004)}]{krapivsky-amjphys-2004}
\bibinfo{author}{\bibfnamefont{P.~L.} \bibnamefont{Krapivsky}}
  \bibnamefont{and} \bibinfo{author}{\bibfnamefont{S.}~\bibnamefont{Redner}},
  \bibinfo{journal}{Am.\ J.\ Phys.} \textbf{\bibinfo{volume}{72}},
  \bibinfo{pages}{591} (\bibinfo{year}{2004}).

\bibitem[{\citenamefont{Rador}(2006)}]{rador-pre-2006b}
\bibinfo{author}{\bibfnamefont{T.}~\bibnamefont{Rador}},
  \bibinfo{journal}{Phys.\ Rev.\ E} \textbf{\bibinfo{volume}{74}},
  \bibinfo{pages}{051105} (\bibinfo{year}{2006}).

\bibitem[{\citenamefont{Serino and Redner}(2010)}]{serino-jstatmech-2010}
\bibinfo{author}{\bibfnamefont{C.~A.} \bibnamefont{Serino}} \bibnamefont{and}
  \bibinfo{author}{\bibfnamefont{S.}~\bibnamefont{Redner}},
  \bibinfo{journal}{J. Stat. Mech.} p. \bibinfo{pages}{P01006}
  (\bibinfo{year}{2010}).

\bibitem[{\citenamefont{Herrmann}(2010)}]{herrmann-prd-2010}
\bibinfo{author}{\bibfnamefont{J.}~\bibnamefont{Herrmann}},
  \bibinfo{journal}{Phys.\ Rev.\ D} \textbf{\bibinfo{volume}{82}},
  \bibinfo{pages}{024026} (\bibinfo{year}{2010}).

\bibitem[{\citenamefont{Ali and Grosskinsky}(2010)}]{ali-acs-2010}
\bibinfo{author}{\bibfnamefont{A.}~\bibnamefont{Ali}} \bibnamefont{and}
  \bibinfo{author}{\bibfnamefont{S.}~\bibnamefont{Grosskinsky}},
  \bibinfo{journal}{Adv.\ Complex Syst.} \textbf{\bibinfo{volume}{13}},
  \bibinfo{pages}{349} (\bibinfo{year}{2010}).

\bibitem[{\citenamefont{Ferrari et~al.}(2006)\citenamefont{Ferrari, Martin, and
  Pimentel}}]{ferrari-pre-2006}
\bibinfo{author}{\bibfnamefont{P.~A.} \bibnamefont{Ferrari}},
  \bibinfo{author}{\bibfnamefont{J.~B.} \bibnamefont{Martin}},
  \bibnamefont{and} \bibinfo{author}{\bibfnamefont{L.~P.~R.}
  \bibnamefont{Pimentel}}, \bibinfo{journal}{Phys.\ Rev.\ E}
  \textbf{\bibinfo{volume}{73}}, \bibinfo{pages}{031602}
  (\bibinfo{year}{2006}).

\bibitem[{\citenamefont{Ali et~al.}(2012)\citenamefont{Ali, Somfai, and
  Grosskinsky}}]{ali-pre-2012}
\bibinfo{author}{\bibfnamefont{A.}~\bibnamefont{Ali}},
  \bibinfo{author}{\bibfnamefont{E.}~\bibnamefont{Somfai}}, \bibnamefont{and}
  \bibinfo{author}{\bibfnamefont{S.}~\bibnamefont{Grosskinsky}},
  \bibinfo{journal}{Phys.\ Rev.\ E} \textbf{\bibinfo{volume}{85}},
  \bibinfo{pages}{021923} (\bibinfo{year}{2012}).

\bibitem[{\citenamefont{Ali et~al.}()\citenamefont{Ali, Somfai, Ball, and
  Grosskinsky}}]{followup}
\bibinfo{author}{\bibfnamefont{A.}~\bibnamefont{Ali}},
  \bibinfo{author}{\bibfnamefont{E.}~\bibnamefont{Somfai}},
  \bibinfo{author}{\bibfnamefont{R.~C.} \bibnamefont{Ball}}, \bibnamefont{and}
  \bibinfo{author}{\bibfnamefont{S.}~\bibnamefont{Grosskinsky}},
  \bibinfo{howpublished}{in preparation}.

\bibitem[{\citenamefont{Abe and Okamoto}(2001)}]{tsallis}
\bibinfo{author}{\bibfnamefont{S.}~\bibnamefont{Abe}} \bibnamefont{and}
  \bibinfo{author}{\bibfnamefont{Y.}~\bibnamefont{Okamoto}},
  \emph{\bibinfo{title}{Nonextensive Statistical Mechanics and its
  Applications}} (\bibinfo{publisher}{Springer-Verlag},
  \bibinfo{address}{Heidelberg}, \bibinfo{year}{2001}).

\bibitem[{\citenamefont{Lebovka and
  Vygornitskii}(1998)}]{lebovka-jphysamathgen-1998}
\bibinfo{author}{\bibfnamefont{N.~I.} \bibnamefont{Lebovka}} \bibnamefont{and}
  \bibinfo{author}{\bibfnamefont{N.~V.} \bibnamefont{Vygornitskii}},
  \bibinfo{journal}{J.\ Phys.\ A: Math.\ Gen.} \textbf{\bibinfo{volume}{31}},
  \bibinfo{pages}{9199} (\bibinfo{year}{1998}).

\bibitem[{\citenamefont{Shlesinger et~al.}(1995)\citenamefont{Shlesinger,
  Zaslavsky, and Frisch}}]{levyflights-book-1995}
\bibinfo{editor}{\bibfnamefont{M.~F.} \bibnamefont{Shlesinger}},
  \bibinfo{editor}{\bibfnamefont{G.~M.} \bibnamefont{Zaslavsky}},
  \bibnamefont{and} \bibinfo{editor}{\bibfnamefont{U.}~\bibnamefont{Frisch}},
  eds., \emph{\bibinfo{title}{L\'evy Flights and Related Topics in Physics}}
  (\bibinfo{publisher}{Springer}, \bibinfo{address}{Berlin},
  \bibinfo{year}{1995}).

\bibitem[{\citenamefont{Biagini et~al.}(2010)\citenamefont{Biagini, Hu,
  {\O}ksendal, and Zhang}}]{fbm-book-2010}
\bibinfo{author}{\bibfnamefont{F.}~\bibnamefont{Biagini}},
  \bibinfo{author}{\bibfnamefont{Y.}~\bibnamefont{Hu}},
  \bibinfo{author}{\bibfnamefont{B.}~\bibnamefont{{\O}ksendal}},
  \bibnamefont{and} \bibinfo{author}{\bibfnamefont{T.}~\bibnamefont{Zhang}},
  \emph{\bibinfo{title}{Stochastic Calculus for Fractional Brownian Motion and
  Applications}} (\bibinfo{publisher}{Springer}, \bibinfo{address}{Berlin},
  \bibinfo{year}{2010}).

\bibitem[{\citenamefont{Alemany and ben Avraham}(1995)}]{alemany-1995}
\bibinfo{author}{\bibfnamefont{P.~A.} \bibnamefont{Alemany}} \bibnamefont{and}
  \bibinfo{author}{\bibfnamefont{D.}~\bibnamefont{ben Avraham}},
  \bibinfo{journal}{Phys. Lett. A} \textbf{\bibinfo{volume}{206}},
  \bibinfo{pages}{18 } (\bibinfo{year}{1995}).

\bibitem[{\citenamefont{Munasinghe et~al.}(2006)\citenamefont{Munasinghe,
  Rajesh, Tribe, and Zaboronski}}]{zaboronski-2006}
\bibinfo{author}{\bibfnamefont{R.}~\bibnamefont{Munasinghe}},
  \bibinfo{author}{\bibfnamefont{R.}~\bibnamefont{Rajesh}},
  \bibinfo{author}{\bibfnamefont{R.}~\bibnamefont{Tribe}}, \bibnamefont{and}
  \bibinfo{author}{\bibfnamefont{O.}~\bibnamefont{Zaboronski}},
  \bibinfo{journal}{Comm. Math. Phys} \textbf{\bibinfo{volume}{268}},
  \bibinfo{pages}{717} (\bibinfo{year}{2006}).

\bibitem[{\citenamefont{Kirkpatrick et~al.}(1983)\citenamefont{Kirkpatrick,
  {Gelatt Jr}, and Vecchi}}]{kirkpatrick-science-1983}
\bibinfo{author}{\bibfnamefont{S.}~\bibnamefont{Kirkpatrick}},
  \bibinfo{author}{\bibfnamefont{C.~D.} \bibnamefont{{Gelatt Jr}}},
  \bibnamefont{and} \bibinfo{author}{\bibfnamefont{M.~P.}
  \bibnamefont{Vecchi}}, \bibinfo{journal}{Science}
  \textbf{\bibinfo{volume}{220}}, \bibinfo{pages}{671} (\bibinfo{year}{1983}).

\end{thebibliography}

\end{document}